\documentstyle[preprint,aps,prd,epsf]{revtex}
\tighten

\begin{document}
\draft
\title{Free energy of bubbles and droplets in the quark-hadron phase transition}
\author{Gregers Neergaard and Jes Madsen}
\address{Institute of Physics and Astronomy, University of Aarhus,
DK-8000 {\AA}rhus C, Denmark}
\date{Published in Phys.Rev. D 60, 054011, 1 September 1999-issue} 
\maketitle

\begin{abstract}
Using the MIT bag model, 
we calculate the free energy of droplets of quark-gluon plasma in a
bulk hadronic medium, and of hadronic bubbles in a bulk quark-gluon
plasma, under the assumption of vanishing chemical potentials.
We investigate the validity of the multiple reflection
expansion approximation, and 
we advise a novel procedure for calculating finite-size corrections to
the free energy of hadronic bubbles in a bulk quark-gluon plasma.
While our results agree largely with earlier calculations,
we show that the usual multiple reflection expansion should be used
with caution, and 
we propose a modification of the multiple reflection expansion,
which makes this approximation agree nicely with direct numerical
calculations. The results should be of relevance in connection
with the cosmological
quark-hadron transition as well as for ultrarelativistic heavy ion
collisions.
\end{abstract}

\pacs{12.38.Mh, 12.39.Ba, 98.80.Cq}


\section{INTRODUCTION}
\label{sec.intro}
The quark-hadron phase transition is of significant interest in
connection with ultrarelativistic heavy ion collision experiments, the
interior of neutron stars, and the evolution of the early Universe.
A calculation from first principles using QCD is at present impossible,
but lattice-QCD studies have shed some light on the transition, for
instance demonstrating, that the transition is apparently first order
for pure glue, whereas the order for physical QCD is still a matter
of investigation.

Awaiting more definite answers to come from such investigations,
numerous studies have been performed using phenomenological models, in
order to gain insight into the physics of the transition. Many such
studies have used the MIT bag model, which in a relatively simple manner
incorporates confinement in terms of a set of boundary conditions for
quarks and gluons.

A very interesting result of a detailed study within the MIT bag model 
was presented by Mardor and Svetitsky \cite{mardor}, who considered the
zero chemical potential case of relevance for the cosmological
quark-hadron transition. For a droplet of quark-gluon plasma within
a bulk medium of pions, a direct numerical calculation of the partition
sum using quark and gluon energy levels led to a behavior of free energy
as a function of radius, $F(R)$, as expected for a first order
transition, namely a minimum of $F$ for $R=0$ when $T$ is below the
transition temperature, $T_0$, and an energy barrier for $R$ of order a few fm
separating a local minimum at $R=0$ from the true minimum (diverging
negative energy) for $R\rightarrow \infty$.

To treat the ``inverse'' problem of a 
vacuum (hadron) bubble within a bulk phase
of quark-gluon plasma, the authors employed a phase shift formula to
calculate the changes in quark and gluon density of states stemming from
the presence of the hadron bubble; again calculating the contribution to
the free energy by a direct numerical integration. In this case, a
peculiar feature was observed, namely that $F(R)$ had a negative minimum
for radii of 1--2 fm, even for $T>T_0$, apparently indicating an
instability of the quark-gluon plasma above $T_0$, since there was
no energy barrier to prevent formation of hadron bubbles.

An interpretation of the result was put forward in terms of an expansion
of the free energy in terms of volume, surface, and curvature
contributions,
\begin{equation}
F(R)=\Delta P \frac{4}{3} \pi R^3+\sigma 4\pi R^2+\alpha 8\pi R +...
\label{fintro}
\end{equation}
Here $R$ is the radius of the droplet or bubble, $\Delta P$ is the pressure
difference between quark and hadron phases (with $\Delta P=0$ defining
the transition temperature, $T_0$), $\sigma$ is the surface tension, and
$\alpha$ the curvature coefficient, where volume, surface and curvature
terms can be calculated from the smoothed quark and gluon densities of state
within the MIT bag model (see below). The results of Ref.\ \cite{mardor}
were apparently well reproduced under the assumption that a vacuum
bubble behaves like a plasma droplet turned inside out, so that the
radius changes sign. This leaves the area term unchanged, but volume and
curvature terms change sign. Or, if $R$ is defined to be positive in
the expression above, then $\sigma$ is unchanged, but pressure
difference as well as curvature coefficient changes sign when going from
the case of a plasma droplet to that of a vacuum bubble. The fact that
the curvature contributions from massless quarks and gluons are very
large compared to the surface contributions coming only from the massive
s-quarks, could explain that an energy barrier for the plasma droplet
could turn into an energy minimum in the reverse case of a hadron bubble.

The authors of Ref.~\cite{mardor} were careful to point out a number
of reasons to be cautious about the result. First of all that the MIT
bag model is clearly just a phenomenological model, and also that the
radii of relevance for the interesting bubbles/droplets were perhaps too
small to justify the ideal gas approximations. But if the result was of
a physical nature, it did have important implications for the
understanding of the quark-hadron transition \cite{mardor,lana,kaja,chri}. 
And the procedure of
$R\rightarrow -R$ gave a simple recipe for treating other situations,
such as quark-hadron mixed phases in neutron stars. Some consequences,
however, were rather strange. For instance, it apparently pays
energetically to fill a strangelet with vacuum bubbles, so that it looks
more like a Swiss cheese than like a uniform mixture of quarks
\cite{michael}.

The aim of the present paper is to compare several different
calculations of the free energy of a vacuum bubble embedded in quark-gluon
plasma as well as a quark-gluon plasma droplet within a bulk phase of hadrons.
For the plasma droplet we focus on a direct sum over states compared with
a multiple reflection expansion, showing that terms beyond volume, surface,
and curvature in the free energy are necessary to avoid unphysical 
behavior for small radii. We demonstrate how the next important contributions
to $F$ (proportional to $T\ln (RT)$, and to $T$) arise naturally if
the density of states is truncated below some value of $kR$ (where $k$
denotes momentum) instead of integrating over unphysical, negative 
values for the density of states all the way from $k=0$. 
For the vacuum bubble we confirm the
results of the phase shift approach of Ref.\ \cite{mardor} by comparing to
a more direct sum over states approach introduced below, which we refer
to as the {\em concentric spheres method}. Again we show how 
an improvement of the multiple
reflection expansion leads to correction terms in $F$, such terms arising
naturally from a truncation of the density of states.

The general framework and basic equations are described 
in Section~\ref{sec.theo}. In Section~\ref{sec.numres} 
we present our numerical results,
largely confirming the calculations of Ref.~\cite{mardor}.
Our results show how and why the usual version of the multiple reflection
expansion is not always accurate.
In Section~\ref{sec.corrmre} we show how further terms in the analytical
expansion of the free energy proposed in the literature improves the
agreement with the numerical results, and we show how a physically
motivated truncation of the density of states from the multiple
reflection expansion resolves most of the problems 
encountered in Section~\ref{sec.numres}.
Section~\ref{sec.concl} contains our conclusions.

\section{THEORETICAL FRAMEWORK}
\label{sec.theo}
In this section, we give the basic equations needed for an
analysis of the quark-hadron phase transition within the MIT bag model.
We consider the case of zero chemical potential which is of particular relevance to 
the cosmological quark-hadron transition, but also of interest for
ultrarelativistic heavy ion collisions.
The quark-gluon plasma is taken to consist of three quark flavors (u,d and s),
the corresponding antiquarks,
and eight non-interacting gluons, these particles being described by the
MIT bag model presented below. The hadron phase 
is considered a mixture of the 3 pions $\pi^0,\pi^\pm$,
since all other (much heavier) hadrons contribute only insignificantly
to the free energy.
Further, we shall assume that the pions contribute volume terms only 
(see Sec.~\ref{sec.thermrel}).
We have taken $m_u=m_d=0$, $m_s=150$~MeV and $m_{\pi}=138$~MeV.

\subsection{The MIT bag model}
\label{sec.mitbag}
The MIT bag model \cite{bagmodel1,bagmodel2} is defined by the Lagrangian 
\begin{equation}
L = \int_{\Omega} d^3x\,({\cal L}_{QCD}-B).
\end{equation}
${\cal L}_{QCD}$ is the usual QCD Lagrangian density,
and $L=0$ outside the bag. $\Omega$ is the bag-volume.
$B>0$ is a phenomenological parameter, the bag constant,
which models the difference in energy density between the perturbative
vacuum inside the bag and the non-perturbative QCD 
vacuum outside the bag.
Requiring the action $W=\int_{t_1}^{t_2} dt\, L$
to be stationary with respect to variations of the fields
yields the equations of motion.

At the surface of the bag, the fields are taken to 
satisfy boundary conditions
which correspond to the fields being confined inside the bag-volume.
We neglect gluon-exchange interactions.

The equations of motion for the fields become
the Dirac equation for the quark fields, and the source-free
Maxwell equations for the gluon fields.
The complete set of equations governing the behavior of the fields, 
including the boundary conditions, is 
\begin{equation}
(i\gamma^\mu \partial_\mu -m)\Psi(x) = 0,\ \ \vec{x} \in \Omega
\label{diraceq}
\end{equation}
\begin{equation}
\partial_\mu F^{\mu\nu}(x) = 0,\ \ \vec{x} \in \Omega
\label{maxwelleq}
\end{equation}
\begin{equation}
i n_\mu \gamma^\mu \Psi(x) = \Psi(x),\ \ \vec{x} \in \partial\Omega 
\label{boundq}
\end{equation}
\begin{equation}
n_\mu F^{\mu\nu}(x) = 0,\ \ \vec{x} \in \partial\Omega
\label{boundg}
\end{equation}
in the following notation:
$x=(x^0,\vec{x})$ is a space-time four-vector,
$\partial\Omega$ is the surface of the bag-volume $\Omega$,
for $\vec{x}\in\partial\Omega$ we define
$n^\mu(x)=(0,-\vec{x}/|\vec{x}|)$ as an 
inward-directed unit-normal three-vector to the surface of the bag,
$\Psi(x)$ is the quark-spinor (there will be one for each 
quarkflavor (u,d,s,...) and one for each of the three color
states of a quark) and
$F^{\mu\nu}(x)=\partial^\mu A^\nu(x) -\partial^\nu A^\mu(x)$ 
is the (non-interacting) gluon field (there are eight copies
of this field).

We fix the bag constant by demanding bulk pressure 
balance at the transition temperature.
Somewhat symbolically the bag constant is thus determined by the equation
\begin{equation}
B=\lim_{V \rightarrow \infty}\left\{
-\frac{\partial F_{quarks}}{\partial V}-\frac{\partial F_{gluons}}{\partial V}+
\frac{\partial F_{pions}}{\partial V}\right\}_{T=T_0} ,
\label{bdet}
\end{equation}
$F$ being the free energy.
In the following, we shall set the transition temperature 
to $T_0=150$~MeV, thus fixing the bag constant 
$B=312.6$~MeV/fm$^3=(221.4$~MeV$)^4$.

We can immediately write down the expression for the gluon field,
since this is just the solution to the source-free Maxwell
equations. 
Expressing the gluon field in terms of color-electric 
and color-magnetic fields, writing $A^\mu(x)=(V(x),\vec{A}(x))$, 
$\nabla\times\vec{A}=\vec{B}(x)$ and 
$-\nabla V(x)-\frac{\partial \vec{A}(x)}{\partial x^0}=\vec{E}(x)$,
there are two sets of solutions to (\ref{maxwelleq}), 
labeled TM and TE
(the $l=0$ fields are absent since for $l=0$,
the only solution to the source-free Maxwell equations
is $\vec{B}_{00}=\vec{E}_{00}=0$ \cite{jackson}):
\begin{equation}
\left\{\vec{B}^{TM}_{lm}e^{-x^\mu k_\mu},\vec{E}^{TM}_{lm}e^{-x^\mu k_\mu}
\right\}^{l=1,2,..}_{m=-l,-l+1,...,l},
\label{maxwellsoltm}
\end{equation}
\begin{equation}
\left\{\vec{B}^{TE}_{lm}e^{-x^\mu k_\mu},\vec{E}^{TE}_{lm}e^{-x^\mu k_\mu}  
\right\}^{l=1,2,..}_{m=-l,-l+1,...,l}, 
\label{maxwellsolte}
\end{equation}
where
\begin{eqnarray}
\vec{B}^{TM}_{lm}
&=& -i\,f_l(kr)\,\vec{x}\times(\nabla Y_{lm}(\theta,\phi)) \\
&=& -i\,\nabla\times(\vec{x}\,f_l(kr)\,Y_{lm}(\theta,\phi)), \label{btm}
\end{eqnarray}
\begin{eqnarray}
\vec{E}^{TM}_{lm} 
&=& \frac{i}{k}\,\nabla\times\vec{B}^{TM}_{lm}(x) \\
&=& \frac{1}{k}\nabla\times\nabla\times(\vec{x}\,f_l(kr)\,Y_{lm}
(\theta,\phi)), \label{etm}
\end{eqnarray}
\begin{eqnarray}
\vec{E}^{TE}_{lm} 
&=& -i\,f_l(kr)\,\vec{x}\times(\nabla Y_{lm}(\theta,\phi)) \\
&=& -i\,\nabla\times(\vec{x}\,f_l(kr)\,Y_{lm}(\theta,\phi)), \label{ete}
\end{eqnarray}
\begin{eqnarray}
\vec{B}^{TE}_{lm} 
&=& -\frac{i}{k}\,\nabla\times\vec{E}^{TE}_{lm}(x) \\
&=& -\frac{1}{k}\nabla\times\nabla\times(\vec{x}\,f_l(kr)\,Y_{lm}
(\theta,\phi)), \label{bte}
\end{eqnarray}
$Y_{lm}(\theta,\phi)$ are the usual spherical harmonics,
$f_l(z)=a\,j_l(z)+b\,n_l(z)$,
the spherical Bessel-functions $j_l(z)$ and $n_l(z)$ 
being the two linearly independent solutions of the equation
$$z^2g''(z)+2zg'(z)+(z^2-l(l+1))g(z)=0,$$
$r=|\vec{x}|$, and $k^\mu=(|\vec{k}|,\vec{k})$.
The constants $a$ and $b$ appearing in the function 
$f_l$, and the possible values of $k$, must
be fixed from the boundary conditions (\ref{boundg}).
Expressed in terms of the fields $\vec{E}$ and $\vec{B}$,
these boundary conditions read:
\begin{equation}
\vec{x}\cdot\vec{E} = \vec{x}\times\vec{B} = 0,\ \ \vec{x} \in \partial \Omega.
\label{boundgeb}
\end{equation}

\begin{figure}
\epsfxsize=8.6truecm
\epsfbox{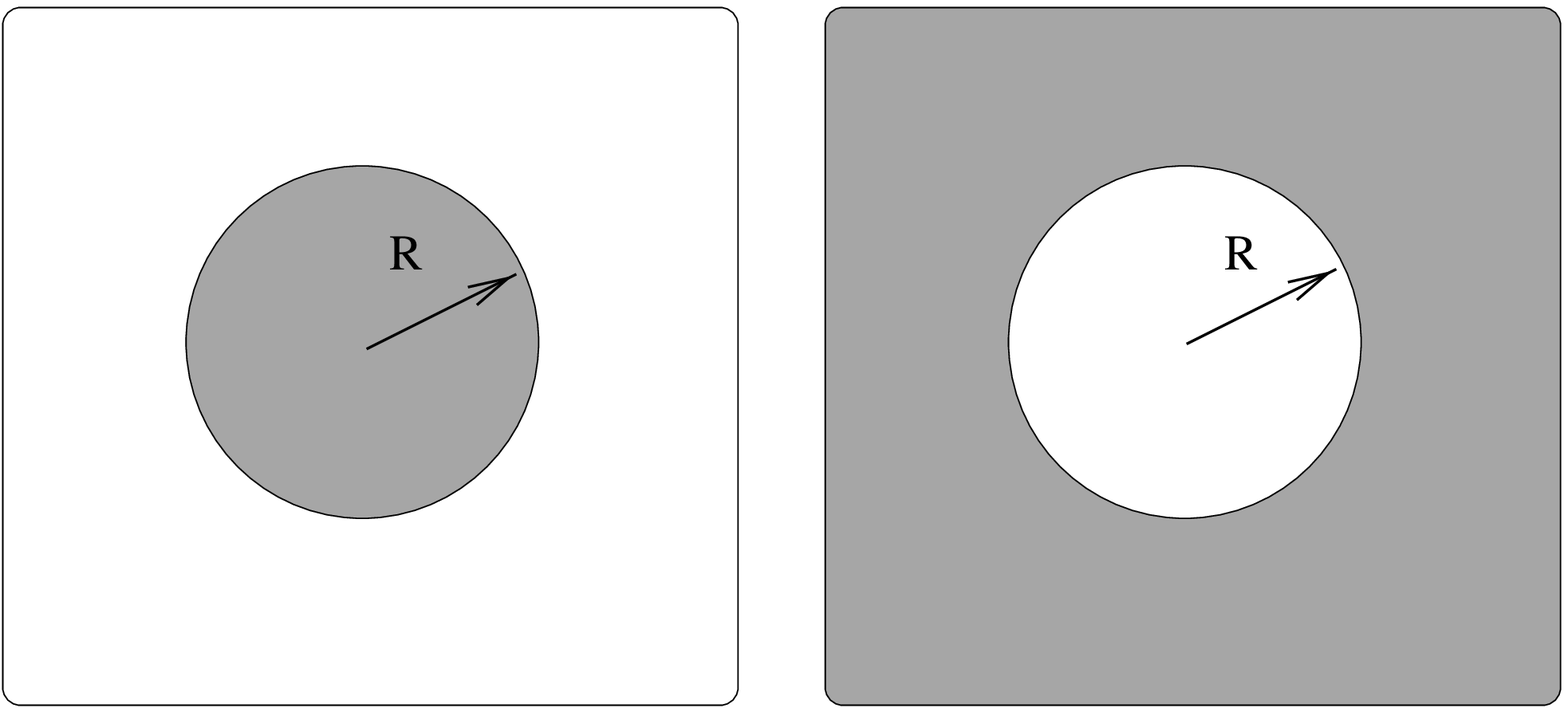}
\caption[]{
Left: A plasma droplet with non-perturbative vacuum outside; this
is essentially an MIT bag.
Right: A vacuum bubble surrounded by plasma, 
boundary conditions being those of the MIT bag, but
corresponding to no flux of plasma {\em into} the bubble.
The phase outside a droplet/bubble extends to infinity.}
\label{fig:plavac}
\end{figure}

\paragraph{Extension of the MIT bag model.}
The MIT bag is a finite region of space(-time) to which quarks and
gluons are confined by boundary conditions~(\ref{boundq})--(\ref{boundg})
corresponding to no flux of plasma out of the droplet. 
We shall refer to this configuration as a plasma droplet.
However, in the following we shall also use the MIT bag model in a slightly
different way, namely the case where quarks and gluons are kept 
{\em outside} a finite region of space, cf.\ Fig.~\ref{fig:plavac}.
The equations describing the quarks and gluons in the second configuration, 
the ``vacuum bubble'', are still~(\ref{diraceq})--(\ref{boundg}), 
but now using $n^\mu(x)=(0,\vec{x}/|\vec{x}|)$ in the boundary conditions.

\subsection{Thermodynamical relations}
\label{sec.thermrel}
For a system of non-interacting fermions (upper sign)
or bosons (lower sign) we can calculate the free energy
for each particle degree of freedom as
\begin{equation}
F(T,V) = \mp T \sum_{i=1}^{\infty} \ln (1 \pm e^{-E_i(V)/T}) 
\label{Fni}
\end{equation}
where $E_i(V)=\sqrt{m^2+k_i^2(V)}$.  
In the continuum case we have
\begin{equation}
F(T,V) = 
\mp T \int d^3k\, \tilde\rho(\vec{k},V) \ln (1 \pm e^{-\sqrt{m^2+k^2}/T}) ,
\label{fnicont}
\end{equation}
$\tilde\rho(\vec{k},V)$ being the density of states, 
defined such that $\tilde\rho(\vec{k},V)d^3k$ is the number of 
states in the volume $V$ with momentum in $d^3k$ around $\vec{k}$.
The importance of the free energy stems from the fact that
the configuration realized in nature is characterized by
a minimum in this free energy.

In the following we shall speak of the volume part resp.\ the surface 
part of the free energy. 
In the case of non-interacting Dirac 
particles and non-interacting gluons 
(these are just Maxwell fields) which are the particle species
relevant to us, the density of
states $\tilde\rho$ in any sufficiently large volume 
$V$ contains a term proportional
to the volume. 
In fact\footnote{
$g_i$ accounts for spin (helicity) degeneracy.}, 
\begin{equation}
\tilde\rho(V\!\rightarrow\!\infty) \simeq g_i\frac{V}{8\pi^3}
\label{rhov}
\end{equation}
independent of which particle species we consider. 
Therefore, also the free energy will contain a term proportional to the volume
of the system. We name this term the {\em volume free energy}.
The total free energy being $F_{tot}$, we can write
$F_{tot}=f_{vol}\,V + F_{sur}$, 
where $F_{sur}/V \rightarrow 0$ as $V\!\rightarrow\!\infty$, 
and $f_{vol}$ does not depend on the volume. 
We shall call $F_{sur}$ the surface part of the free energy,
or simply the {\em surface free energy}.

\subsection{The multiple reflection expansion, MRE}
\label{sec.mre}
The multiple reflection expansion (MRE) is an approximation for the
density of states, also commonly referred to as 
the asymptotic expansion of the density of states.
Since we only consider systems with spherical symmetry, 
we define the spherically symmetric density of states
$\rho(k,V) \equiv 4\pi k^2 \tilde\rho(\vec{k},V)$. 
Consider a spherical volume $V=\frac{4\pi}{3}R^3$ of quarks and gluons, 
described by the bag model.
The MRE for this system, 
as a sum of volume, area, and curvature contributions,
valid for sufficiently large volumes, is
\begin{equation}
\rho_i(k,V) = 
\frac{Vk^2}{2\pi^2}+f_{A,i}(k/m)\,k\,4\pi R^2+f_{C,i}(k/m)8\pi R+...,
\ \ i=q,g
\label{rho}
\end{equation}
where
\begin{equation}
f_{A,q}(k/m) = -\frac{1}{8\pi}(1-\frac{2}{\pi}\arctan(k/m)) 
\label{rhofaq}
\end{equation}
\begin{equation}
f_{C,q}(k/m) = \frac{1}{12\pi^2}(1-\frac{3k}{2m}(\frac{\pi}{2}-\arctan(k/m)))
\label{rhofcq}
\end{equation}
\begin{equation}
f_{A,g} = 0
\label{rhofag}
\end{equation}
\begin{equation}
f_{C,g} = -\frac{1}{6\pi^2}. 
\label{rhofcg}
\end{equation}
Here, $A$ stands for area, $C$ for curvature,
and indices $q$ and $g$ denote quarks and gluons.
Note that 
$\lim_{m \rightarrow 0} f_{C,q}(k/m) = -\frac{1}{24\pi^2}$,
and that
$\lim_{m \rightarrow 0} f_{A,q}(k/m) = 0$.

The MRE was developed by Balian and Bloch~\cite{bb1},
and the above expressions for the area- and curvature-terms
have appeared in the literature.
The area term for quarks is given (though not derived) 
in \cite{berger}.
The curvature term for massless quarks 
seems to appear explicitly for the first time 
in \cite{farhi}, whereas the full 
expression (\ref{rhofcq}) for massive quarks 
is introduced in \cite{madsen}.
The gluon expressions, valid for non-interacting gluons, 
is calculated in \cite{bb2}.

As indicated by the dots in~(\ref{rho}), the expression for 
$\rho_i(k,V)$ should in principle contain terms proportional
to $1/R$, $1/R^2$ etc., but 
as these terms become small in the limit of large $R$, and since
the MRE is an approximation valid for large systems, these terms
are usually neglected. However, we shall see in the following 
that the MRE as it stands in~(\ref{rho}) is not only inaccurate, 
but also unphysical at small radii, having negative density of states.
Further, we shall argue that, when used in calculations of the free energy,
the MRE~(\ref{rho}) containing only area- and curvature-terms
leads to errors even at larger radii, where the MRE
itself {\em is} a good approximation to the density of states.
We also suggest a solution to these problems.

Everywhere in the following, unless explicitly stated,
reference to the MRE means the approximation~(\ref{rho}) 
to the density of states {\em without} further correction 
terms like $1/R$, $1/R^2$.

\subsection{The inverse multiple reflection expansion, MRE($-R$)}
\label{sec.invmre}
Because of the ``symmetry'' between the two situations
(i) quark-gluon plasma confined by MIT bag boundary conditions 
within a sphere of radius $R$ (a ``plasma droplet''), 
and (ii) quark-gluon plasma kept outside a sphere of radius $R$ 
by MIT bag boundary conditions (a ``vacuum bubble''),
it has been argued~\cite{mardor} that
there should exist a simple relation between the density of states
in the two cases, i.e.\ that 
the density of states of quarks and gluons in the case of a vacuum
bubble can be found from the expressions~(\ref{rho})--(\ref{rhofcg})
for a plasma bubble by simply inverting the sign of $R$.
We shall refer to this hypothesis as the MRE($-R$).
Since the MRE($-R$) is derived from the MRE, we 
expect the MRE($-R$) to have problems related to those of the 
MRE mentioned in the previous subsection.

\subsection{The phase shift approach}
In this subsection we briefly describe the phase shift approach 
to calculating the free energy of a vacuum bubble.
The phase shift formula~(\ref{phs}) 
was introduced in this context by Mardor and Svetitsky~\cite{mardor}.
The phase shift approach is based on a relation between the density of states
and the scattering phase shifts\footnote{
In the case of spherical symmetry,
the phase shifts $\delta_l(k)$ are defined such that 
the effect of the scatterer
is to change the spatial part of the wave function far away from the scatterer
for a given angular
momentum $l$ from $\propto\frac{1}{kr}\sin(kr-l\pi/2)$ to 
$\propto\frac{1}{kr}\sin(kr-l\pi/2+\delta_l(k))$.}:
\begin{equation}
\Delta\rho_l(k) = \frac{1}{\pi} \frac{d\delta_l(k)}{dk}. 
\label{phs}
\end{equation}
For a derivation of this relation in the non-relativistic case, 
see e.g.~\cite{huang}.
Here, $\Delta\rho_l$ is the change in the density of states (at a given
angular momentum) induced by the scatterer.

In order to use this phase shift approach to calculate the free energy, 
we need the scattering phase shifts for quarks
and gluons. These are derived from the defining 
equations~(\ref{diraceq}) and~(\ref{boundq}) for the quarks, and
from~(\ref{maxwellsoltm}), (\ref{maxwellsolte}) and (\ref{boundgeb})
for the gluons.
The phase shift for the $j$-component of the quark field
($j=1/2,3/2,...$ is total angular momentum)
is the sum of two components
\begin{equation}
\delta_j(k) = \delta_j^{l=j-1/2}(k) + \delta_j^{l=j+1/2}(k),
\label{deltaqtot}
\end{equation}
where, for a surface of radius $R$,
\begin{equation}
\delta_j^{l=j\mp 1/2}(k) = 
\arctan \left( \frac{j_l(kR) \pm 
\frac{k}{E+m_q}j_{l\pm 1}(kR)}{n_l(kR) \pm 
\frac{k}{E+m_q}n_{l\pm 1}(kR)}\right),
\label{deltaq}
\end{equation}
and $E=\sqrt{m_q^2+k^2}$, $q=u,d,s.$
The phase shift for the gluon field also consists of two parts, 
$\delta_l^{TM}(k)$ and $\delta_l^{TE}(k)$, where
\begin{eqnarray}
\delta_{l}^{TE}(k) 
&=& \arctan \left( \frac{\frac{d}{dr}(rj_l(kr))}
  {\frac{d}{dr}(rn_l(kr))} \right)_{r=R} \\
&=& \arctan \left( \frac{j_l(kR)(1+l)-kRj_{l+1}(kR)}
  {n_l(kR)(1+l)-kRn_{l+1}(kR)} \right), \label{deltatm}
\end{eqnarray}
and
\begin{equation}
\delta_{l}^{TM}(k) = \arctan \left( \frac{j_l(kR)}{n_l(kR)} \right),
\label{deltate}
\end{equation}
again for a surface of radius $R$.
$l=1,2,...$ labels orbital angular momentum, 
and here, as opposed to the quark situation, 
it is a good quantum number.

Knowing the phase shifts, the contribution to the free energy 
from the quarks and gluons outside a vacuum bubble of radius $R$ is 
calculated using~(\ref{fnicont}), so that
\begin{equation}
F_i(T,R) = 
  \mp g_i\frac{T}{\pi}\int_{0}^{\infty} dk\, \frac{d\delta_i(k,R)}{dk} 
  \ln(1 \pm e^{-E(k)/T})
\label{phaseshiftform}
\end{equation}
The label $i$ stands for different particle types 
(quarks, gluons) {\em and} angular momentum. 
Again, the upper sign applies
to the fermions (quarks), lower sign to bosons (gluons).
The appropriate degeneracy factors are $g_{quark}=6$ and $g_{gluon}=8$. 

We make two remarks about the formulae~(\ref{phs}) 
and~(\ref{phaseshiftform}).
(i)~The free energy~(\ref{phaseshiftform}) includes the contribution 
from the excluded volume.
(ii)~When the phase shifts contain functions with multiple branches, 
like the $\arctan$ function in our case, we choose the branch
which makes the phase shifts continuous functions of the energy.
(In the case of potential scattering where the potential
obeys certain integrability conditions, one can prove that the
phase shifts are continuous functions of the energy~\cite{amrein}.)

\subsubsection{The free energy in the limit $RT \rightarrow 0$ }
By expanding the Bessel-functions 
appearing in~(\ref{deltaq}),~(\ref{deltatm}) and~(\ref{deltate}) as
\begin{equation}
j_l(x) = \sum_{k=0}^{\infty} a_k(l)\,x^{l+2k} 
\label{jbesexpan}
\end{equation}
and
\begin{equation}
n_l(x) = \sum_{k=0}^{\infty} b_k(l)\,x^{2k-l-1}, 
\label{nbesexpan}
\end{equation}
(valid for $l>0$) and keeping only the lowest order terms,
we obtain via~(\ref{phaseshiftform}) 
the following analytical expressions
valid for $RT \ll 1$ for the surface 
free energy of (one flavor of) massless quarks
(index $j$ and $l$ means that we consider each
angular momentum component separately)
\begin{equation}
\frac{F_{S,q}^j(RT)}{T} \simeq 
-12\,\frac{(2j+1)\,(2^{2j+2}-1)\,\pi^{2j+2}}{(2j+3)}\,
\alpha_q(j)\,(RT)^{2j+2} B_{j+3/2}
\label{fqvaclim}
\end{equation}
and for the eight gluons
\begin{equation}
\frac{F_{S,g}^l(RT)}{T} \simeq
-8 \frac{(2l+1)(2\pi)^{2l+1}}{2l+2}\,\alpha_g(l)\,(RT)^{2l+1}\,B_{l+1} 
\label{fgvaclim}
\end{equation}
where 
\begin{equation}
\alpha_q(j) = 
\frac{a_0(j+1/2)\,b_0(j+1/2)-a_0(j-1/2)\,b_0(j-1/2)}{b_0^2(j+1/2)} ,
\label{alfaq}
\end{equation}
\begin{equation}
\alpha_g(l) = 
\frac{a_0(l)}{b_0(l)}+\frac{a_0(l)\,(l+1)}{b_0(l)\,(l+1)-b_0(l+1)} 
\label{alfag}
\end{equation}
and
\begin{equation}
a_k(l) = 
\frac{(-1)^k}{2^k\,k!\,1\cdot 3\cdot 5 \cdots (2l+2k+1)} ,
\label{akl}
\end{equation}
\begin{equation}
b_k(l) = 
(-1)^{k+1}\,\frac{1\cdot 3\cdot 5\cdots (2l-1)}
{2^k\,k!\,(1-2l)(3-2l)\cdots (2k-1-2l)} .
\label{bkl}
\end{equation}
(The factors $(1-2l)(3-2l)...$ in the denominator of~(\ref{bkl})
appear only when $2k-1 \geq 1$).
The $B_{n}$ appearing in~(\ref{fqvaclim}) and~(\ref{fgvaclim})
are the Bernoulli numbers, defined by
\begin{equation}
\frac{x}{e^x-1} = 
1-\frac{x}{2}+B_1\,\frac{x^2}{2!}-B_2\,\frac{x^4}{4!}+B_3\,\frac{x^6}{6!}-...
\label{bernnodef}
\end{equation}
the first few of these being
\begin{equation}
B_1=1/6,\ \ B_2=1/30,\ \ B_3=1/42,\ \ B_4=1/30.
\label{bernnoex}
\end{equation}
On the basis of~(\ref{fqvaclim}) and~(\ref{fgvaclim}),
we conclude that the first energy term of importance for $R\rightarrow 0$
is proportional to $R^3$; no terms proportional to $R$ or $R^2$ 
appear in this limit.  This is in contrast to the MRE($-R$) conjecture,
where a curvature term proportional to $R$ dominates for $R\rightarrow 0$.
The difference is clearly demonstrated in the figures in the next section
as a difference between zero and finite slope of the free energy 
for $R\rightarrow 0$.

\subsection{The concentric spheres method}
\begin{figure}
\epsfxsize=4.0truecm
\epsfbox{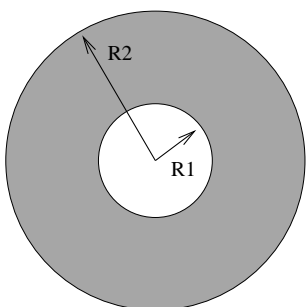}
\caption[]{
The concentric spheres configuration: 
The inner sphere has radius $R_1$, the outer sphere has radius $R_2$.
There is non-perturbative vacuum inside the inner sphere 
and outside the outer sphere.
Quark-gluon plasma is confined between the two spheres by
MIT bag boundary conditions corresponding to no flux of plasma 
across the spheres.
At the outer sphere the boundary conditions are the usual MIT bag
conditions~(\ref{boundq})--(\ref{boundg}), but at the inner sphere
we use $n^\mu(x)|_{r=R_1} = -n^\mu(x)|_{r=R_2} = (0,\vec{x}/|\vec{x}|).$}
\label{fig:consph}
\end{figure}

\begin{figure}
\epsfxsize=8.6truecm
\epsfbox{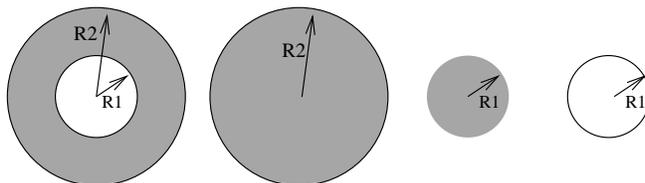}
\caption[]{
How to extract the surface free energy in the concentric spheres method:
Shaded areas are plasma, white areas are non-perturbative vacuum.
Contributions to free energy (left to right):
$F_{total}(R_1,R_2)$, $F_{total}(R_2)$, $F_{volume}(R_1)$
and $F_{surface}(R_1)$.
$F_{total}(R_1,R_2)$ and $F_{total}(R_2)$ are calculated 
by summation over energy levels of the particles 
in the relevant configuration.
$F_{volume}(R_1)$ is only the {\em volume} free energy
of particles occupying a volume $\frac{4\pi}{3}R_1^3$.
The purpose is to calculate the contribution to the free energy
from the inner surface, $F_{surface}(R_1)$, and this can be done
as in~(\ref{fsurcs}).}
\label{fig:consph2}
\end{figure}

The concentric spheres method is a new way to calculate the surface
contribution to the free energy of 
quarks and gluons outside a vacuum bubble.
The idea is to extract the contribution from the inner surface
to the total free energy of
the concentric spheres configuration in Fig.~\ref{fig:consph}.
Assuming that the splitting of the free energy in volume and surface
contributions is valid, we can extract the free energy contribution 
from the inner surface (cf. Fig.~\ref{fig:consph2}) 
from a calculation of the total
free energy of particles contained between two concentric spheres as
\begin{equation}
F_{surface}(R_1)=F_{total}(R_1,R_2)+F_{volume}(R_1)-F_{total}(R_2),
\label{fsurcs}
\end{equation}
where
$F_{total}(R_1,R_2)$ is the total free energy (including both surface contributions)
of the particles contained between the spheres with radii $R_1$ and $R_2$,
$F_{volume}(R_1)$ is the volume contribution to the free energy of
the particles in a sphere of radius $R_1$, and
$F_{total}(R_2)$ is the total free energy (including surface contribution) 
of the particles in a sphere of radius $R_2$.
It is precisely $F_{surface}(R_1)$, the surface free energy of the particles
outside a sphere, that we are interested in.
Here, the terms $F_{total}(R_1,R_2)$ and $F_{total}(R_2)$ are calculated
directly by summation over energy levels, whereas $F_{volume}(R_1)$ is
calculated from the MRE (using a positive radius).
Explicitly, the term $F_{total}(R_1,R_2)$ in the case of quarks 
of mass $m_q$ is
\begin{equation}
F_{quarks}(R_1,R_2)=-6T\sum_{j=1/2,3/2}^{\infty}(2j+1)\sum_{l=j\pm 1/2}
\sum_{n} \ln(1+e^{-E_{jln}/T}),\ \ E_{jln}=\sqrt{k_{jln}^2+m_q^2},
\label{fcsq}
\end{equation}
where $k_{j,l=j\pm 1/2,n}$ is the $n$'th solution of the equation
\begin{eqnarray}
(\alpha(k) j_{l-1}(kR_1) - j_l(kR_1)) \,
(\alpha(k) n_{l-1}(kR_2) + n_l(kR_2)) - \nonumber \\
(\alpha(k) j_{l-1}(kR_2) + j_l(kR_2)) \,
(\alpha(k) n_{l-1}(kR_1) - n_l(kR_1)) = 0
\label{cs_ljplus}
\end{eqnarray}
for $l=j+1/2$, and 
\begin{eqnarray}
(\alpha(k) n_{l+1}(kR_1) + n_l(kR_1)) \,
(j_l(kR_2) - \alpha(k) j_{l+1}(kR_2)) + \nonumber \\
(\alpha(k) n_{l+1}(kR_2) - n_l(kR_2)) \,
(\alpha(k) j_{l+1}(kR_1) + j_l(kR_1)) = 0
\label{cs_ljminus}
\end{eqnarray}
for $l=j-1/2$, with 
$\alpha(k)=\frac{k}{\sqrt{k^2+m_q^2}\,+m_q}$.
In the case of gluons, the term $F_{total}(R_1,R_2)$ is
\begin{equation}
F_{gluons}(R_1,R_2)=8T\sum_{l=1}^{\infty}(2l+1)\sum_{a=TM,TE}
\sum_{n} \ln(1-e^{-k^a_{l,n}/T}),
\label{fcsg}
\end{equation}
where $k^a_{l,n}$ is the $n$'th solution of the equation
\begin{equation}
j_l(kR_2)\, n_l(kR_1) - j_l(kR_1)\, n_l(kR_2) = 0
\label{cs_tm}
\end{equation}
for the TM gluons, and
\begin{eqnarray}
(l+1)^2\,\{j_l(kR_1)\,n_l(kR_2)-j_l(kR_2)\,n_l(kR_1)\} +\nonumber\\
(l+1)\,kR_1\,\{j_l(kR_2)\,n_{l+1}(kR_1)-j_{l+1}(kR_1)\,n_l(kR_2)\} +\nonumber\\
(l+1)\,kR_2\,\{j_{l+1}(kR_2)\,n_l(kR_1)-j_l(kR_1)\,n_{l+1}(kR_2)\} +\nonumber\\
kR_1\,kR_2\,\{j_{l+1}(kR_1)\,n_{l+1}(kR_2)-j_{l+1}(kR_2)\,n_{l+1}(kR_1)\} = 0,
\label{cs_te}
\end{eqnarray}
for the TE gluons.
The term $F_{total}(R_2)$ is just the free energy of 
a quark- or gluon-droplet, so this is calculated 
using Eqs.~(\ref{qenergies})--(\ref{fgsumoverstates}).
Finally, the volume term $F_{volume}(R_1)$ is
\begin{equation}
F_{volume}(R_1) = 
\mp g_i T \int_0^{\infty} dk\, \frac{Vk^2}{2\pi^2} 
\ln(1 \pm e^{-E(k)/T}),\ \ i=q,g
\label{fcs_vol}
\end{equation}
where $V=\frac{4\pi}{3}R_1^3$, $g_q=12$ for each flavor, and $g_g=16$.

Formally, the splitting in surface and volume terms is appropriate
only when $R_2 -R_1 \rightarrow \infty$.
However, calculations for 10~fm$<R_2<20$~fm 
suggest that the concentric spheres method yields the correct free
energy contribution from the inner surface
up to $R_1 \simeq R_2/2$.

\section{NUMERICAL RESULTS}
\label{sec.numres}
We are now going to use the different techniques described in the 
previous section to calculate the free energy of (i)~a plasma droplet
in a bulk hadronic medium, and of (ii)~a hadron bubble in a bulk plasma.
In both cases we normalize 
the free energy such that it is zero when there is no droplet resp.\ bubble, 
i.e.\ we calculate the free energy
relative to an infinite hadron resp.\ plasma phase.
In the adopted model, we have the following contributions 
to the free energy: Quarks (u, d, and s, and their anti-quarks), gluons,
the bag contribution $BV_{QGP}$, where $V_{QGP}$
is the plasma volume, and
the contribution from the three pions (using~(\ref{fnicont})
with $\rho_\pi(k,V)=3\frac{V\,k^2}{2\pi^2}$ and $m=m_\pi$, this is
\begin{equation}
F_{\pi}(T,V_\pi) = \frac{3TV_\pi}{2\pi^2} 
\int_0^\infty dk\,k^2\,\ln(1-e^{-\sqrt{m_\pi^2+k^2}/T}),
\label{fpion}
\end{equation}
where $V_\pi$ is the pion volume).
The bag and pion contributions are thus simple and universal,
the interesting part of the free energy is the quark and gluon contributions.

\subsection{Free energy of a plasma droplet in a bulk hadronic medium}
The plasma droplet is an MIT bag with pions around it, so the calculation
of the free energy is straightforward, i.e.\ we can calculate the energy
levels of quarks and gluons in the bag, and perform the partition 
sum~(\ref{Fni}) directly. This, of course, gives the true free energy.

We can also use the MRE~(\ref{rho}) to calculate the free 
energy~(\ref{fnicont}).
This way, two approximations are involved: (1)~The spectrum is discrete,
but the MRE treats the energy levels as continuous; 
and (2)~we have discarded terms of the form 
$1/R$, $1/R^2$ etc.\ in~(\ref{rho}).
Since we are considering the high temperature case, the first
approximation is well justified: The low ($E<T$)
energy levels, where the level spacing is large, 
do not contribute significantly to the free energy,
whereas the main contribution to the true free energy~(\ref{Fni}) 
comes from the higher levels where the spacing is small.
Thus, any difference in the free energy between a direct calculation
and the MRE approximation is a measure of the importance 
of the neglected terms in~(\ref{rho}) and/or the choice of truncation
discussed in Sec.~\ref{sec.corrmre}.

\subsubsection{Direct calculation}
In this case, we need to solve the set of 
equations~(\ref{diraceq})--(\ref{boundg}), and then perform the 
sum~(\ref{Fni}) over these levels.
We thus obtain the following equations
(to be solved numerically) for the quarks ($l=j\pm 1/2$)
\begin{equation}
j_l(kR) = \pm \frac{k}{E+m_q}\,j_{l\pm 1}(kR),\ \ E=\sqrt{k^2+m_q^2}.
\label{qenergies}
\end{equation}
For the TM-gluons
\begin{equation}
j_l(kR) = 0,
\label{tmgenergyeq}
\end{equation}
and for the TE-gluons
\begin{equation}
j_l(kR)(l+1) = kR\,j_{l+1}(kR).
\label{tegenergyeq}
\end{equation}
These equations provide a series of solutions, that we label
$E_{jln}$ for the quarks and $k^a_{ln},\ a=TM,TE$
for the gluons.
The contribution to the free energy from quarks and 
gluons are then
\begin{equation}
F_q(T,R) = 
  -6T\sum_{j=1/2,3/2,..}^{\infty}(2j+1)\sum_{l=j\pm 1/2}\sum_{n}
  \ln(1+e^{-E_{jln}/T}) 
\label{fqsumoverstates}
\end{equation}
and
\begin{equation}
F_g(T,R) = 
  8T\sum_{l=1}^{\infty}(2l+1)\sum_{a=TM,TE}\sum_{n}
  \ln(1-e^{{-k^a_{ln}}/T}) 
\label{fgsumoverstates}
\end{equation}
respectively.
Note that when dealing with each angular momentum component
separately, the degeneracy factors $g_i$
only account for the degeneracy due to color and particle/anti-particle, 
so we have $g_q=6$ for each quark flavor and $g_g=8$ for the gluons.

We imagine a spherical plasma droplet of volume $V_{QGP}=\frac{4\pi}{3}R^3$
embedded in a large volume of pions,
the total volume of this system being $V_\infty$.
The pions therefore inhabit a volume 
$V_\pi=V_\infty-V_{QGP}$. 
Since we calculate the free energy relative to a system
with no plasma droplet, and pions in the whole volume $V_\infty$,
the effective pion volume in Eq.~(\ref{fpion}) is $-V_{QGP}$.
Summing all the contributions we obtain Fig.~\ref{fig:fpla}.
In Fig.~\ref{fig:fplaop} we show the different contributions 
to Fig.~\ref{fig:fpla}.
We shall comment on these figures later.

\subsubsection{Using the MRE}
Now let us see how the MRE approximation handles the plasma droplet.
The difference from the sum over states method lies entirely
in the calculation of the quark and gluon contributions.
The pion and bag contributions are the same as before.
Now, we use~(\ref{fnicont}) for the quarks and the gluons 
with the MRE density of states~(\ref{rho}).
The result is shown in Fig.~\ref{fig:fas}.
Figure~\ref{fig:fasop} shows the different contributions for $T=152$~MeV.

\subsubsection{Comparison}
\begin{figure}
\epsfxsize=8.6truecm
\epsfbox{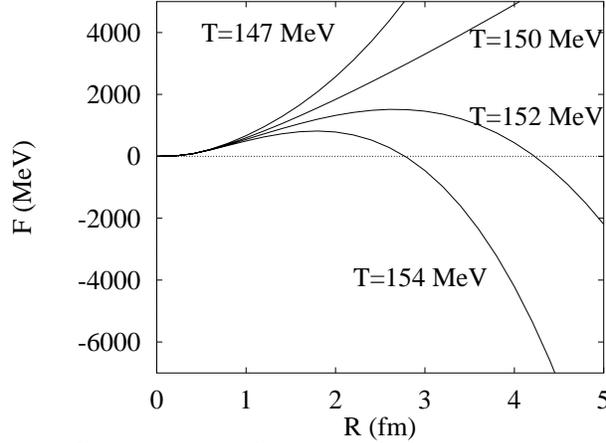}
\caption[]{
Total free energy (calculated directly by summation over states)
of a quark-gluon plasma droplet of radius $R$ 
surrounded by pions. The phase transition temperature is set to 
$T_0=150$~MeV. Results are shown for several temperatures around $T_0$.}
\label{fig:fpla}
\end{figure}

\begin{figure}
\epsfxsize=8.6truecm
\epsfbox{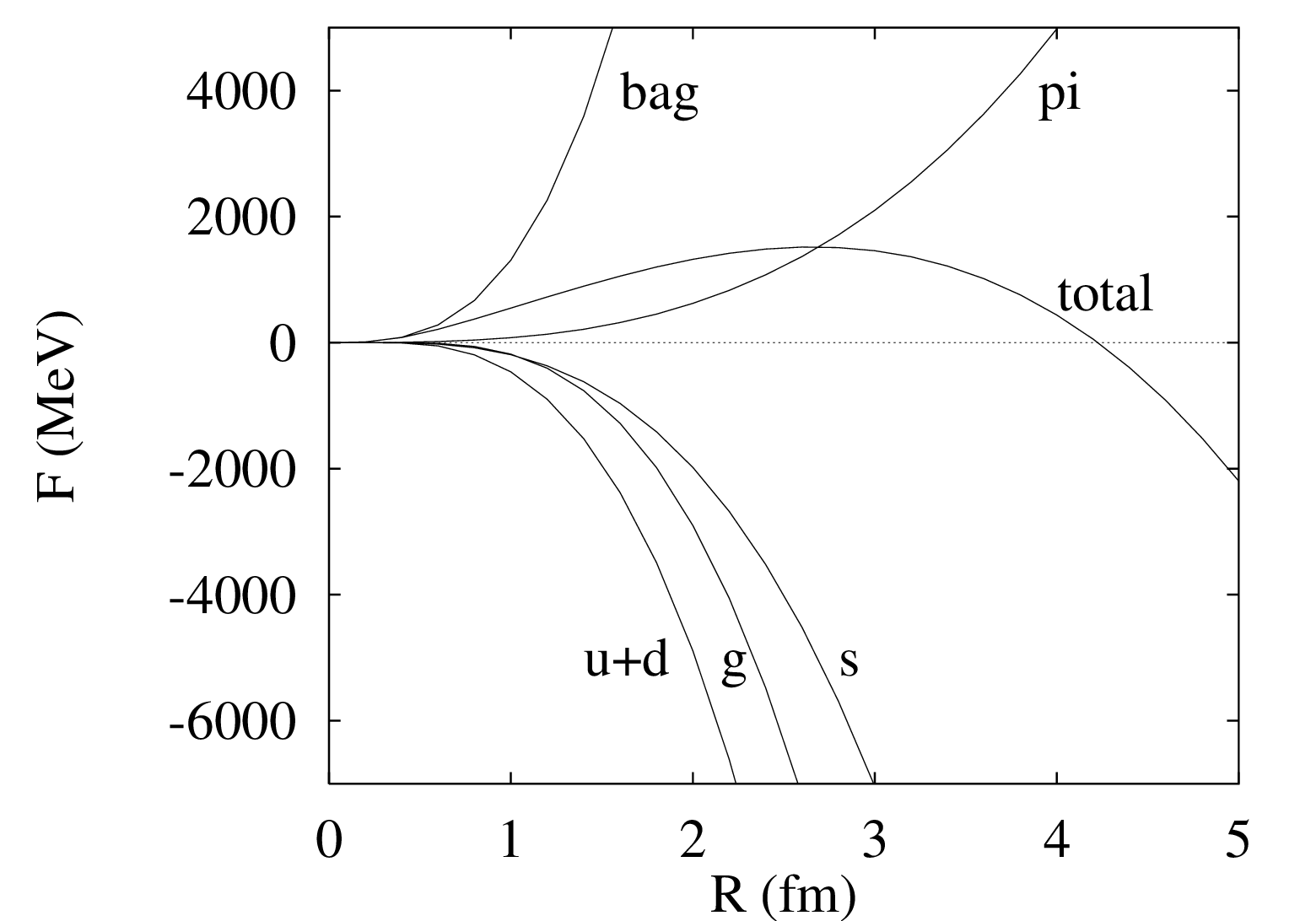}
\caption[]{
Different contributions to Fig.~\ref{fig:fpla} for T=152~MeV.}
\label{fig:fplaop}
\end{figure}

\begin{figure}
\epsfxsize=8.6truecm
\epsfbox{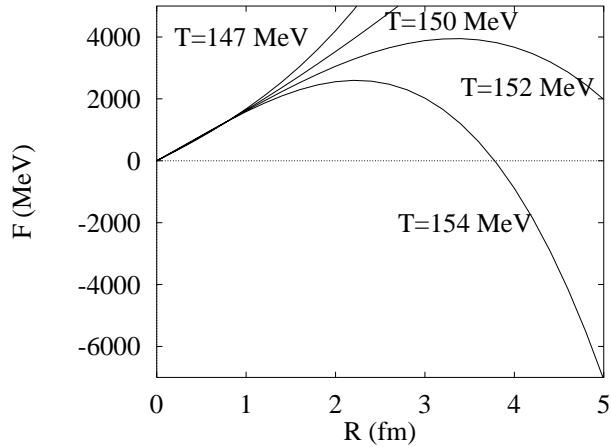}
\caption[]{
As Fig.~\ref{fig:fpla}, except that the free energies are calculated 
using the MRE approximation.}
\label{fig:fas}
\end{figure}

\begin{figure}
\epsfxsize=8.6truecm
\epsfbox{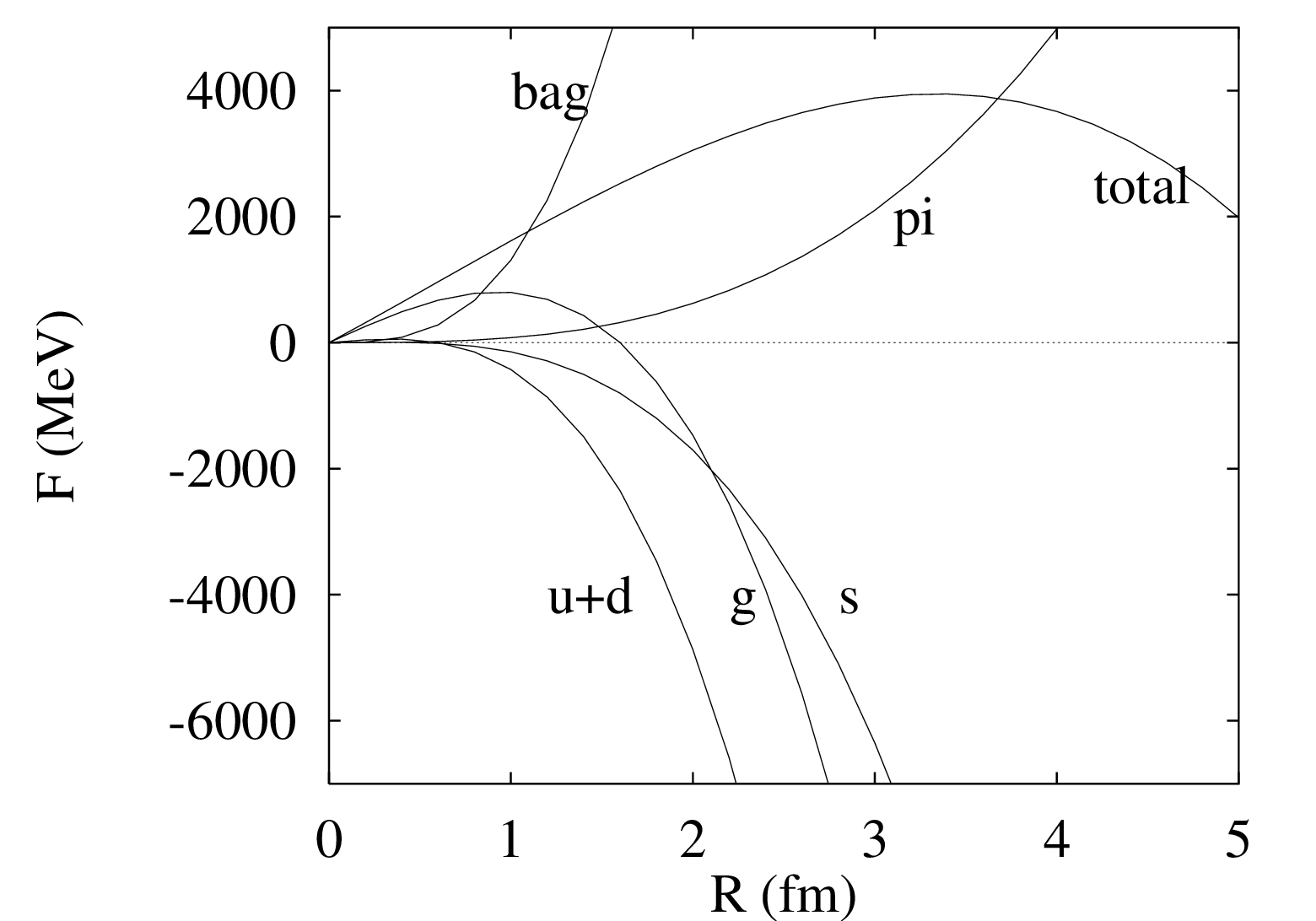}
\caption[]{
The different contributions to Fig.~\ref{fig:fas} for $T=152$~MeV.}
\label{fig:fasop}
\end{figure}

Comparing Fig.~\ref{fig:fpla} and Fig.~\ref{fig:fas} we see that
although they agree qualitatively (except for $R\rightarrow 0$, where
the MRE is dominated by a curvature 
term proportional to $R$, whereas the sum over states for massless particles
behaves as $R^3$), there are significant quantitative
differences even at large radii. At first sight this is surprising. 
The MRE should be
a good approximation, since each of the terms in~(\ref{rho}) is derived
analytically (albeit in the limit $kR \gg 1$). But the MRE is an approximation
for the density of states, not for the free energy itself. Because the
free energy is an integral over the density of states, it ``picks up''
the wrong behavior of the MRE at low energies, and ``remembers''
this error even at larger radii. This is why the free
energy calculated using the MRE is not in quantitative agreement
with the correct free energy in Fig.~\ref{fig:fpla}, although
the MRE is a good approximation for the density of states
in the limit $kR \gg 1$.

Looking at Fig.~\ref{fig:fasop}, we see that the main difference 
between the sum over states approach and the MRE approximation
is due to the gluons. 
The gluon free energy being positive for small ($R<1.5$~fm) radii,
corresponds to the density of states being negative.
This can also be seen directly from~(\ref{rho}) and~(\ref{rhofcg}).
Hence the gluon density of states in the MRE
is not only wrong, it is unphysical at small radii.
This behavior is due to the 
way the MRE handles the surface corrections, namely
through~(\ref{rhofaq})--(\ref{rhofcg}).
Taking more terms ($\propto 1/R$, $\propto 1/R^2$, etc.) 
into account in~(\ref{rho}) might cure this. We propose another
resolution of the problem in Sec.~\ref{sec.corrmre}.

A few words about the physics implied by Figs.~\ref{fig:fpla} 
and~\ref{fig:fas}.
There are two minima, at $R=0$ and at $R=\infty$, separated by
an energy barrier. The picture is therefore the following:
When the temperature is $T<T_0$, no stable droplets can form.
Even when $T>T_0$ there is an energy barrier to pass before stable droplets
of quark-gluon plasma can exist within the pion phase. But once a droplet is 
created (from fluctuations) with radius $R>2$~fm, it will expand unimpeded.
The energy barrier, of course, is due to the surface terms.
The most important quantitative difference between Fig.~\ref{fig:fpla} 
and Fig.~\ref{fig:fas} is the height of the energy barrier
separating the two minima at $R=0$ and $R=\infty$.
The height of the barrier is related to the nucleation rate, 
in this case the plasma formation rate when 
heating a hadron gas e.g.\ in a heavy ion collision. 
Thus, although generally small, the surface contribution
to the free energy has important implications.

\subsection{Free energy of a hadron bubble in bulk plasma}
This configuration is a vacuum bubble with hadrons 
(i.e.\ pions, in our model) inside, surrounded by plasma.
In order to emphasize the essential points, 
we start by focusing on the surface free energy of quarks
and gluons, since this is the interesting, non-universal, 
part of the free energy.

\subsubsection{Surface free energy of a vacuum bubble}
We shall compare results for this surface free energy as calculated 
by the MRE($-R$) conjecture and by the concentric spheres method.
We shall also compare with results obtained by the phase shift
approach. 

\begin{figure}
\epsfxsize=8.6truecm
\epsfbox{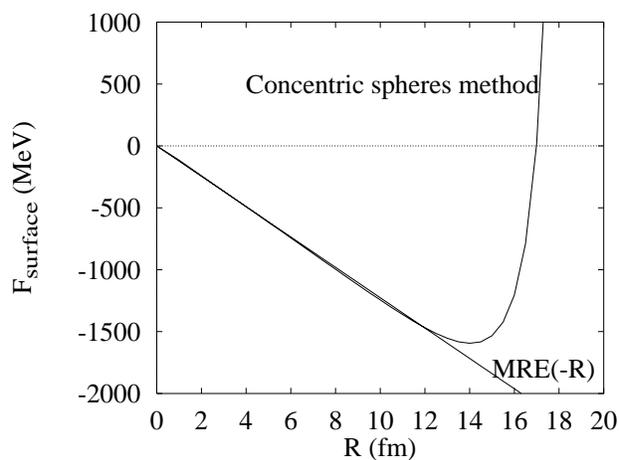}
\caption[]{
A comparison between two different methods of calculating the surface free
energy of massless quarks outside a sphere (a ``vacuum bubble'') 
of radius $R$: 
(1)~Multiple reflection expansion with the 
sign of $R$ reversed (MRE($-R$)), and (2)~the concentric spheres method
with an outer radius $R_2=20$~fm.
When $R<R_2/2$ the two methods yield similar results. This suggests
two things: (A)~When $R\ll R_2$ the interactions at the 
outer surface are unimportant, and
(B)~MRE($-R$) describes adequately the way the inner surface 
alters the density of states in the case of massless quarks.
The temperature is $T=152$~MeV.}
\label{fig:fcs.u}
\end{figure}

\begin{figure}
\epsfxsize=8.6truecm
\epsfbox{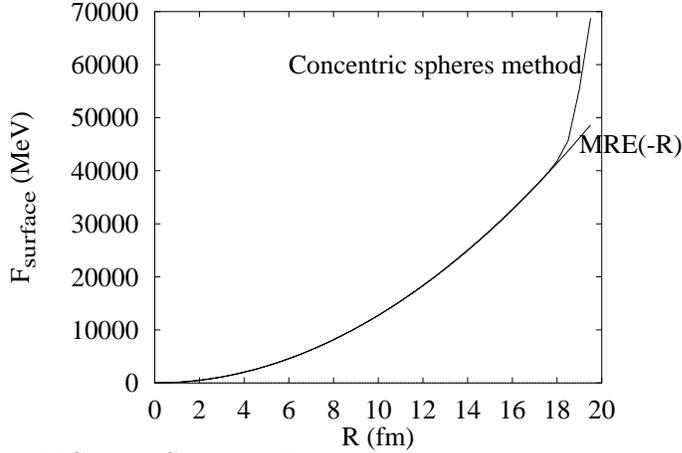}
\caption[]{
Same as Fig.~\ref{fig:fcs.u}, but using a quark mass of 150~MeV.
In this case, the effect of the outer surface is not visible
until $R$ is quite close to the outer surface at $R_2=20$~fm.
Again, the MRE($-R$) seems to be a satisfactory description.}
\label{fig:fcs.s}
\end{figure}

\begin{figure}
\epsfxsize=8.6truecm
\epsfbox{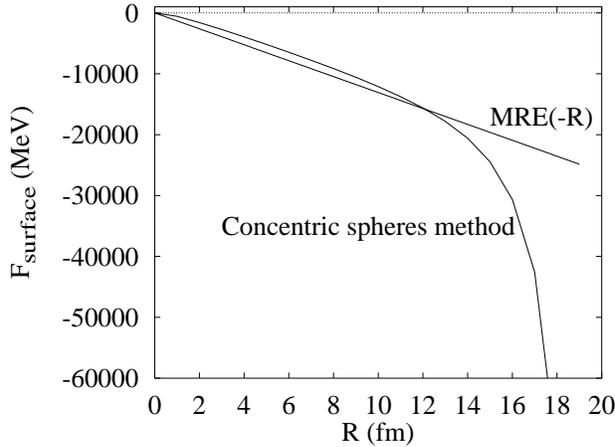}
\caption[]{
Same as Fig.~\ref{fig:fcs.u}, but for gluons.
The MRE($-R$) is not as good a description as in the case of the quarks.}
\label{fig:fcs.g}
\end{figure}

\begin{figure}
\epsfxsize=8.6truecm
\epsfbox{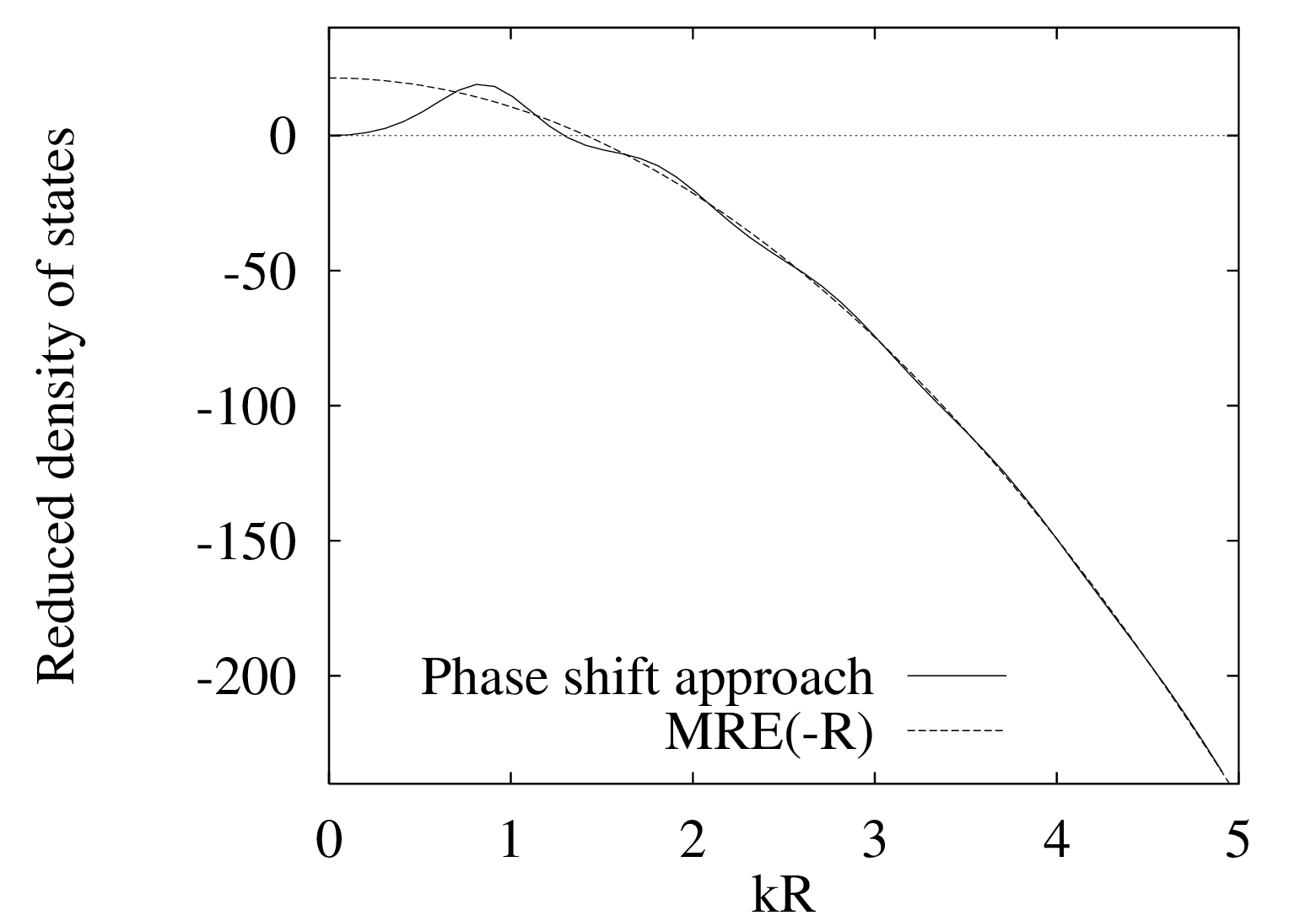}
\caption[]{
The ``reduced density of states'' $\frac{\pi}{R}\rho(kR)$ of gluons
outside a vacuum bubble,
calculated by the phase shift approach, and using the MRE($-R$).
The phase shift method yields the more correct result 
(cf. Fig.~\ref{fig:fcomp}).
Note that this is the density of states relative to a situation with no
vacuum bubble, so $\rho<0$ in this case is not unphysical.} 
\label{fig:rhocomp}
\end{figure}

\begin{figure}
\epsfxsize=8.6truecm
\epsfbox{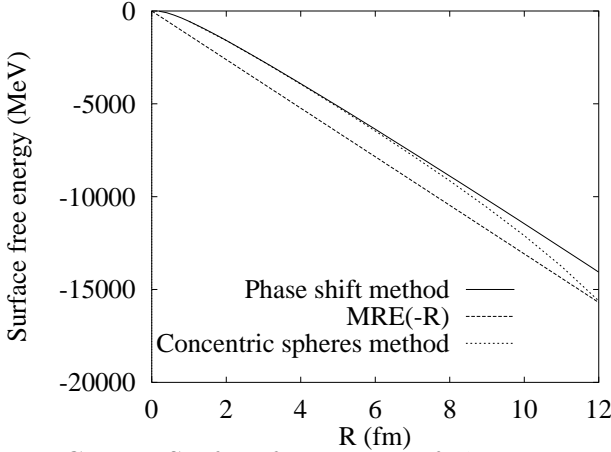}
\caption[]{
Surface free energy of gluons outside a vacuum bubble of radius $R$
calculated in the three different ways described in the text.
The outer radius used in the concentric spheres method is $R_2=20$~fm.
The phase shift approach agrees with the concentric spheres method 
for $R<6$~fm, whereas
the deviation between these two methods at $R>6$~fm
is due to the influence of the outer surface on the result of
concentric spheres method.
To emphasize the differences between the three methods,
we have subtracted the volume free energy from 
the phase shift results, so that only the surface 
contributions are shown in this figure. The temperature is $T=152$~MeV.}
\label{fig:fcomp}
\end{figure}

The figures~\ref{fig:fcs.u}, \ref{fig:fcs.s} and \ref{fig:fcs.g}
show the surface free energy of massless quarks, massive quarks
and gluons respectively. 
We show the results for just one temperature, $T=152$~MeV, 
but the picture is qualitatively the same for other temperatures
($T=50,100,160$~MeV).
These figures show that the MRE($-R$) works quite well in the case
of quarks (massless and massive), but less well for gluons.
The {\em slope} of the gluon MRE($-R$) curve is correct, but there
is an offset of about $1000$~MeV (cf. Fig.~\ref{fig:fcs.g}).
This is because the MRE($-R$) for the density of states, although
a good approximation at large values of $kR$,
is wrong at small $kR$, and since the free energy at a given large radius
is an integral over the density of states also at small momenta,
the bad behavior of the MRE($-R$) at small $kR$
affects the free energy even at
large radii (cf. remarks about the MRE for the plasma droplet). 
The same remarks apply to the quarks, 
but here the effect is much less pronounced.
To support this picture we show in Fig.~\ref{fig:rhocomp}
a comparison of the quantity $\frac{\pi}{R}\rho(kR)$, $\rho$ being
the density of states of the eight gluons, calculated by the MRE($-R$):
$\frac{\pi}{R}\rho_{MRE(-R)}(kR)=-\frac{32}{3}(kR)^2+\frac{64}{3}$,
and by the phase shift approach~(\ref{phaseshiftform}).
In contrast to other figures in this section, Fig.~\ref{fig:rhocomp}
includes the volume contribution.
Figure \ref{fig:fcomp} compares all three methods of calculating 
the surface free energy, here shown in the case of gluons.
We expect the concentric spheres method to be a correct way of
calculating the free energy, as long as we stay in the regime $R \ll R_2$.
The agreement of the concentric spheres method and the phase shift
approach at $R<6$~fm in Fig.~\ref{fig:fcomp} suggests
that both approaches are valid ways to calculate
the surface free energy and/or the density of states of quarks and gluons
outside a vacuum bubble, for the phase shift approach
presumably at any radius.
Referring back to Fig.~\ref{fig:rhocomp} it is therefore clear that
the MRE($-R$) approach is inadequate at small $kR$.
This was to be expected. First, the original MRE (for positive radii) 
is derived in the limit of large $kR$, and
second, we have seen that the MRE is unphysical at small radii,
so it is not surprising that also the MRE($-R$) has problems in this regime.

We thus conclude, that as an approximation of the density of states,
the MRE($-R$) for gluons works well at large values of $kR$, but
is incorrect at small $kR$. 
We have only shown the gluon data, but a similar conclusion 
is valid for the quarks, although the error at small radii is less important
than in the case of gluons.
Further, we have argued that as far as the free energy is concerned, 
we should be even more careful when applying the MRE($-R$),
as the bad behaviour of the MRE($-R$) at small radii manifests itself
as an error in the free energy even at large radii.
Finally, our results obtained with the concentric spheres method
are consistent with the phase shift formula~(\ref{phs}),
which seems to be an accurate way to calculate the density of states
of quark-gluon plasma outside a vacuum bubble.

\subsubsection{The total free energy}

\begin{figure}
\epsfxsize=8.6truecm
\epsfbox{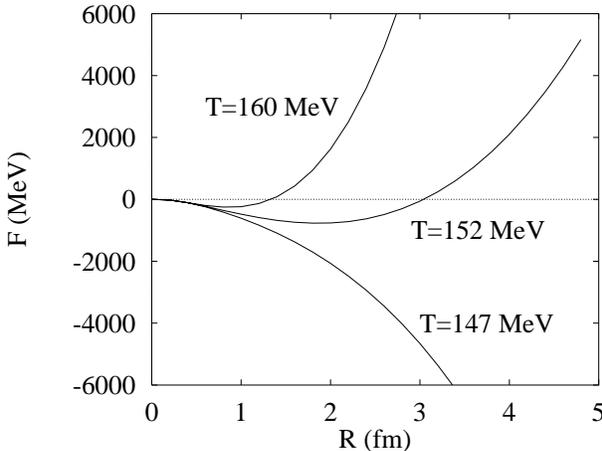}
\caption[]{
Free energy of a pion bubble surrounded by quark-gluon plasma,
normalized so that the free energy of a pure plasma without pions is zero.
Curves are shown for temperatures above and below 
the transition temperature, $T_0=150$~MeV.
The surface contributions from the quarks and gluons are calculated
by the concentric spheres method with an outer surface of $R_2=20$~fm.
The minimum at $R \simeq 1\--2$~fm shows that in this model, 
bubbles of pions of this radius will form even for temperatures
above $T_0$. Similar results were obtained by Mardor and 
Svetitsky~\cite{mardor} using the phase shift method.}
\label{fig:ftotal}
\end{figure}

Knowing the contribution from the surface to the free energy 
of quarks and gluons outside the vacuum bubble, 
we can easily calculate the total free energy of the whole configuration:
The volume contribution of the plasma is calculated using~(\ref{fnicont})
with the smoothed density of states~(\ref{rhov}) inserted. 
The pion contribution is given by~(\ref{fpion}), and the bag constant 
contributes a term $BV_{QGP}$ as always.
Adding these contributions we obtain the results in Fig.~\ref{fig:ftotal}.
The interesting part of this figure is the minimum of the free energy
at $R=1$--$2$~fm for temperatures well above the
transition temperature.
Mardor and Svetitsky~\cite{mardor} found a similar minimum in the
free energy using the same model as described in this paper, but
calculating the free energy of the plasma using the
phase shift approach, whereas here we have applied the concentric spheres
method.

\section{CORRECTIONS TO THE MRE}
\label{sec.corrmre}
We have seen that the MRE and the MRE($-R$) for gluons 
as it stands in~(\ref{rho}), (\ref{rhofag}) and (\ref{rhofcg}) 
have problems at small values of $kR$, leading to errors in the 
free energy even at large radii.
Although we have numerical methods to calculate correctly the free energy
both in the plasma droplet case and in the vacuum bubble case,
we would like to be able to use some MRE approximation to gain
physical insight, and
for practical computations because the direct methods 
are numerically demanding.
In this section, we investigate how to modify the MRE and the MRE($-R$),
in order for these approximations to describe more correctly 
the density of states of the configuration in question.

\subsection{Gluon droplet}
Previously in this paper, we have shown how to calculate exactly 
the free energy of abelian gluons in an MIT bag.
That was, however, a numerical computation.
Analytic calculations of the free energy, not using the MRE, 
have also appeared in the literature.
Using the same model for the gluons as we do, 
De Francia~\cite{defrancia,defranciap} finds for the difference
$\Delta F=F_{\mathrm gluons}-F_{\mathrm gluons,MRE}$
in the limit of large $RT$ 
\begin{equation}
\frac{\Delta F}{T} = -0.874-\frac{5}{8}\ln(RT)+... 
\label{dFdefrancia}
\end{equation}
where the dots indicate terms of higher order in $(RT)^{-1}$.
(De Francia gives such terms explicitly, 
but they are too small to be relevant in our analysis.)
Note that~(\ref{dFdefrancia}) is calculated for one abelian gauge field,
and should thus be multiplied by 8 in order to describe the gluon
free energy.

\begin{figure}
\epsfxsize=8.6truecm
\epsfbox{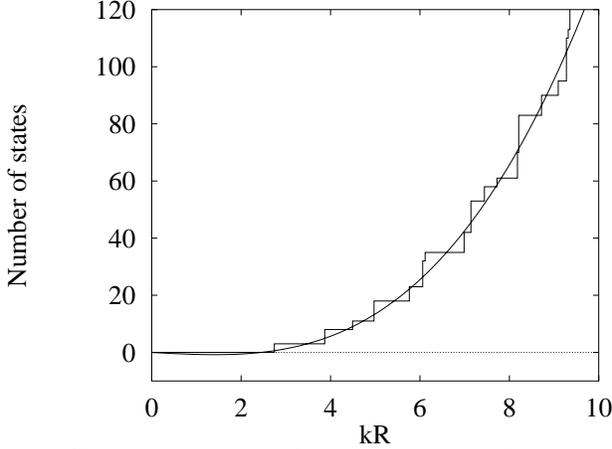}
\caption[]{
Number of gluon states with energy less than $k$ 
in an MIT bag of radius $R$, calculated
(i) directly, solving~(\ref{tmgenergyeq}) and~(\ref{tegenergyeq})  
(discontinuous line, true values) and
(ii) by the MRE (continuous line, approximation).
Note that the MRE predicts a negative density of states
at small values of $kR$.}
\label{fig:nos}
\end{figure}

The main problem with the MRE is that it predicts a negative
density of states at small $kR$ where in reality there are no states,
see Fig.~\ref{fig:nos}.
An error in the density of states at small values of $kR$ 
is particularly severe, since here the statistical factor in the
integrand of the free energy is large.
In the following, we will show that 
using a reduced density of states of the form
\begin{equation} 
\frac{\pi}{R}\rho_{\Lambda}(kR) = \left\{ \begin{array}{ll}
  0, & 0 \leq kR<\Lambda \\
  \frac{2}{3}(kR)^2-\frac{4}{3}, & kR \geq \Lambda
\end{array}
\right. 
\label{mmre}
\end{equation}
cures most of the problems of the MRE.
We shall refer to this density of states, $\rho_{\Lambda}$, 
as the MMRE (modified MRE), since it consists of the usual
MRE-contributions for $kR \geq \Lambda$, but is truncated below
$kR=\Lambda$.
When we are in a regime where $RT \gg 1$, we can find an approximate analytical
expression for the correction 
$\Delta F=F_{\mathrm gluons,MMRE}-F_{\mathrm gluons,MRE}$
to the free energy induced by using 
$\rho_{\Lambda}$ instead of $\rho$ as the density of states of gluons:
\begin{equation}
\frac{\Delta F}{T}=
-16\left(\ln(\Lambda)(\frac{2}{9\pi}\Lambda^3-\frac{4}{3\pi}\Lambda)
-\frac{2}{27\pi}\Lambda^3+\frac{4}{3\pi}\Lambda
+\ln(RT)(\frac{4}{3\pi}\Lambda-\frac{2}{9\pi}\Lambda^3)\right).
\label{dflambda}
\end{equation}
We can fix the value of ${\Lambda}$ by matching the coefficient
of $\ln(RT)$ to the analytical result of De Francia, i.e.\ solving
$16(\frac{2}{9\pi}\Lambda^3-\frac{4}{3\pi}\Lambda)=-5$, which has
$\Lambda=0.832$ as the relevant solution, 
and the free energy~(\ref{dflambda}) becomes
\begin{equation}
\frac{\Delta F}{T} = -6.352-5\ln(RT).
\label{dflambdafix}
\end{equation}
The fact that $6.352 \simeq 8\cdot 0.874$ shows the consistency
of the procedure, cf.\ Eq.~(\ref{dFdefrancia}).

\begin{figure}
\epsfxsize=8.6truecm
\epsfbox{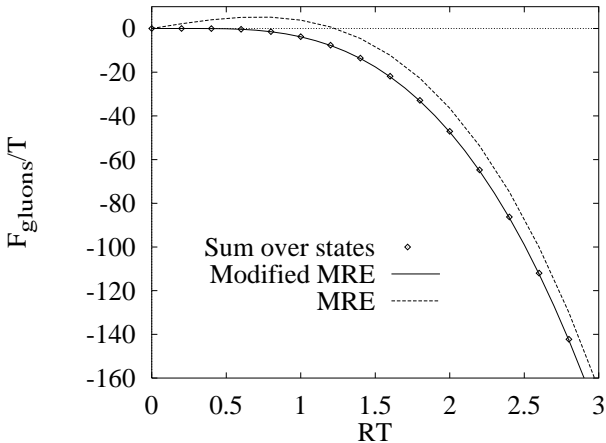}
\caption[]{
Gluon contribution to the free energy (in units of the temperature, $T$)
of a plasma droplet of radius $R$,
as a function of the dimensionless parameter $RT$.
The unphysical behavior of the usual MRE ((\ref{rho}), 
(\ref{rhofag}) and (\ref{rhofcg})) at small radii causes the
free energy to deviate from the true free energy (calculated by
summing over energy levels) even at large radii.
Using the modified MRE, $\rho_{0.832}$, in~(\ref{fnicont})
makes the free energy agree remarkably well with the true free energy.}
\label{fig:ftcomp}
\end{figure}

In Fig.~\ref{fig:ftcomp} we compare the proposal~(\ref{mmre}) 
with a direct calculation 
(summing over energy levels) of the free energy. Also shown is
the free energy calculated using the usual MRE.

\subsection{Vacuum bubble}
Balian and Duplantier~\cite{baldup} have calculated the Casimir energy
of a perfectly conducting spherical shell.
They find (in the large $RT$ limit, and
again not quoting terms of higher order in $(RT)^{-1}$)
\begin{equation}
\frac{\Delta \tilde F}{T} = -\frac{0.769}{4}-\frac{\ln(RT)}{4}.
\label{dFbaldup}
\end{equation}
In our language, $8\cdot\Delta \tilde F$ is the sum of 
(i) the surface free energy of gluons inside an MIT bag, and 
(ii) the surface free energy of gluons outside a vacuum bubble.
Using this and De Francia's calculation~(\ref{dFdefrancia}),
we can deduce the corrections to the MRE($-R$).
We obtain for the difference
$\Delta F_{\mathrm vac}=
F_{\mathrm gluons,corrected}-F_{\mathrm gluons,MRE(-R)}$
\begin{equation}
\Delta F_{\mathrm vac} = T(5.454+3\ln(RT)).
\label{dfvac}
\end{equation}
Like in the gluon droplet case, we can advise a modification of
the MRE($-R$), which works very well. 
Specifically, we propose the following:
The reduced density of states of gluons outside a vacuum bubble, is
\begin{equation} 
\frac{\pi}{R}\rho_{\mathrm vac}(kR) = \left\{ \begin{array}{ll}
  0, & 0 \leq kR<0.458 \\
  -\frac{2}{3}(kR)^2+\frac{4}{3}, & kR \geq 0.458
\end{array}
\right. 
\label{mmre-r}
\end{equation}
where $R>0$ is the radius of the vacuum bubble.
We refer to this density of states as the MMRE($-R$).
The value of $kR=0.458$ where we cut the MRE($-R$) is fixed by the
same procedure as in the gluon droplet case. 
The difference
$\Delta F=F_{\mathrm gluons,MMRE(-R)}-F_{\mathrm gluons,MRE(-R)}$ 
is then 
\begin{equation}
\Delta F = T(5.417+3\ln(RT)),
\label{dfvacfix}
\end{equation}
showing the consistency of the procedure (cf.\ Eq.~(\ref{dfvac})).
The fact that we should cut the density of states at $kR=0.458$ and not
$kR=0.832$ as in the gluon droplet case, reflects the asymmetry
between the gluon droplet and the vacuum bubble configurations.

\begin{figure}
\epsfxsize=8.6truecm
\epsfbox{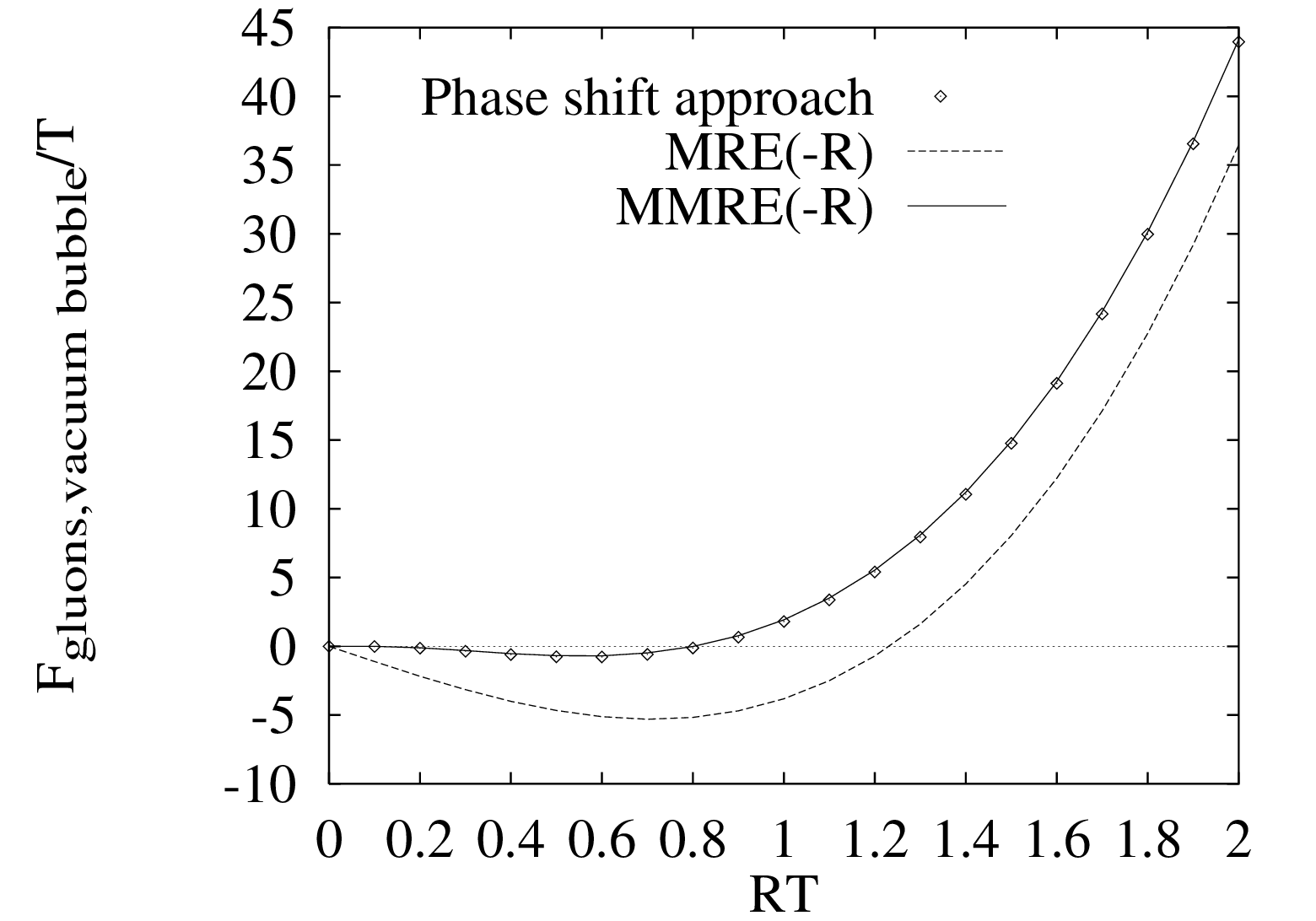}
\caption[]{
The free energy of gluons outside a vacuum bubble of radius $R$
(normalized to the temperature, $T$) calculated in three different ways: 
(1)~By the phase shift approach, which we consider an accurate procedure,
(2)~by the MRE($-R$), and (3)~using the MMRE($-R$).}
\label{fig:ftvac}
\end{figure}

In Fig.~\ref{fig:ftvac} we compare the different methods of 
calculating the free energy of gluons outside a vacuum bubble.
The simple MMRE($-R$) suggestion is in nice 
agreement with the phase shift approach, which (based on our
calculations in the previous sections) we consider the most 
accurate way of calculating the free energy.

\section{CONCLUSION}
\label{sec.concl}
This paper has a twofold purpose. First, we have 
introduced the concentric spheres method as a way to calculate
the free energy of quark-gluon plasma outside a pion bubble,
confirming the peculiar results of Mardor and Svetitsky~\cite{mardor},
that, within the MIT bag model, this free energy has a minimum 
at non-zero radius even well above the transition temperature.

Second, we have shown that terms beyond volume, surface, and curvature
are necessary in order to reproduce the free energy of plasma droplets
and vacuum bubbles within the multiple reflection expansion,
especially for the gluon contributions.
We have discussed the reasons for this, and 
based on previous calculations~\cite{defrancia,baldup},  
we extract correction terms
to the free energy, which can be understood
from a physically motivated truncation of the density of states.

Our calculations are all performed in the limit of vanishing chemical
potentials.
The results are thus relevant to investigations of 
the cosmological quark-hadron transition,
and possibly to forthcoming ultrarelativistic heavy ion collision
experiments at RHIC and LHC.
While these are certainly interesting prospects, we
plan to extend our analysis to situations of finite chemical
potential, such calculations being relevant 
to a wider range of applications including e.g.\ neutron stars.

\acknowledgments
JM was supported in part by the Theoretical Astrophysics Center
under the Danish National Research Foundation, and thanks the Institute
for Nuclear Theory in Seattle, sponsered by DOE for its hospitality. 
We thank Michael
Christiansen for useful discussions, and Ben Svetitsky for important
comments on an earlier version of the manuscript.

\begin{references}
\bibitem{mardor} I. Mardor and B. Svetitsky, Phys. Rev. D {\bf 44}, 878 (1991).
\bibitem{lana} G. Lana and B. Svetitsky, Phys. Lett. B {\bf 285}, 251 (1992).
\bibitem{kaja} K. Kajantie, L. K\"arkk\"ainen and K. Rummukainen, 
Phys. Lett. B {\bf 286}, 125 (1992).
\bibitem{chri} M.B. Christiansen and J. Madsen, J. Phys. G {\bf 23},
2039 (1997).
\bibitem{michael} M.B. Christiansen, PhD-thesis, University of Aarhus (1997).
\bibitem{bagmodel1} A. Chodos, R.L. Jaffe, K. Johnson, C.B. Thorn and
V.F. Weisskopf, Phys. Rev. D {\bf 9}, 3471 (1974).
\bibitem{bagmodel2} K. Johnson, Acta Phys. Pol. {\bf B6}, 865 (1975).
\bibitem{jackson} J. D. Jackson, {\em Classical Electrodynamics} 
(Wiley, New York, 1975), p. 746.
\bibitem{bb1} R. Balian and C. Bloch, Ann. Phys. (N.Y.) {\bf 60}, 401 (1970).
\bibitem{berger} M.S. Berger and R.L. Jaffe, Phys. Rev. C {\bf 35}, 213
(1987); {\it ibid\/} {\bf 44}, 566 (1991) (Erratum).
\bibitem{farhi} E. Farhi and R.L. Jaffe, Phys. Rev. D {\bf 30}, 2379 (1984).
\bibitem{madsen} J. Madsen, Phys. Rev. D {\bf 50}, 3328 (1994).
\bibitem{bb2} R. Balian and C. Bloch, Ann. Phys. (N.Y.) {\bf 64}, 27 (1971);
{\it ibid\/} {\bf 84}, 559 (1974) (Erratum).
\bibitem{huang} K. Huang, {\em Statistical Mechanics}, 2nd ed., 
(Wiley, New York, 1987). 
\bibitem{amrein} W. O. Amrein, J. M. Jauch, and K. B. Sinha,
{\em Scattering Theory in Quantum Mechanics} (Benjamin, Reading, 1977).
\bibitem{defrancia} M. De Francia, Phys. Rev. D {\bf 50}, 2908 (1994).
\bibitem{defranciap} M. De Francia, private communication with Michael Christiansen.
\bibitem{baldup} R. Balian and B. Duplantier, Ann. Phys. {\bf 112}, 165 (1978).
\end{references}
\end{document}